Nano cluster dissociation sensitive to the elasticity of metal as a collision target

Kozo Mochiji, Naoki Se, Norio Inui, Kousuke Moritani
Graduate School of Engineering, University of Hyogo, 2167 Shosha, Himeji, Hyogo 671-2280, Japan

ABSTRACT

The sensitive dependence of the cluster ion dissociation behavior on the elastic response of the target metal is experimentally demonstrated. Five types of metal are bombarded with cluster ions consisting of thousands of argon atoms at an incident kinetic energy per atom of less than 10 eV. A mass spectroscopic analysis of dissociated ions such as $Ar_2^+$ or $Ar_3^+$ is carried out. The branching ratio for the ion yield $Ar_2^+/\Sigma Ar_n^+$ ($n \geq 2$), representing the dissociation rate, is found to be significantly different for each metal. The relationship between the branching ratio and the impulsive stress at the contact between the cluster ion and the metal is investigated. The impulsive stress is calculated based on the Young's modulus of the cluster ion and the metal, under the assumption that the collision is initially elastic. As a result, magnitude correlation in the branching ratio well corresponds with that in the impulsive stress. This result is important in that it suggests a new method to evaluate mechanical property of materials such as Young's modulus by analyzing the collision with nanocluster ions.

A nanoscale cluster containing 100-10,000 atoms or molecules potentially exhibits intermediate property between individual atoms or molecules and bulk materials. Recently, collisions between nanocluster and solid have become important in nanotechnologies such as thin film deposition, powder or granular technology, nanocatalysis, nanomedicine. Interactions between nanoclusters and solids have been studied extensively from a microscopic viewpoint based on a dynamics of the constituents of cluster and solid [1-10]. However, there is little study from a macroscopic viewpoint and it is not clear how much theory or model used in macroscale collisions is available for nanoscale collisions. In the present study, we considered a nanoscale cluster to be a continuum as well as solid, and examined the possibility that classical mechanics could analyze the collision between them.

One important parameter determining the nature of the collision between clusters and solids is the average kinetic energy per constituent atom ($E_{at}$) for the incident cluster. Another is the binding energy ($E_{bi}$) of atoms within the cluster, which is typically comparable to that in the target solid. If $E_{at}$ is significantly higher than $E_{bi}$, the cluster almost completely decomposes and the solid deforms plastically through various physical or chemical reactions. When $E_{at}$ is comparable to $E_{bi}$, the cluster only decomposes partially or remains intact, and the solid deforms in a more elastic manner. We previously examined the deformation of a single graphene sheet due to collisions with clusters consisting of 1000 argon atoms using molecular dynamics simulations [11]. For $E_{at}$ = 5 eV/atom, the graphene sheet was found to vibrate in the cluster beam direction, but not ruptured. Saitoh et al. also simulated induced vibrations in a grapheme sheet due to collisions with clusters of 500 argon atoms at $E_{at}$ = 0.005-0.125 eV/atom. They found that the time evolution of the sheet deflection could be well explained using linear elastic theory [12].

As mentioned above, when $E_{at}$ is comparable to or below $E_{bi}$, most of the kinetic energy of an incident cluster is dissipated through elastic deformation of the target material and dissociation of the cluster. The cluster dissociates into smaller clusters or remains intact, which depends on the magnitude of the impact. Accordingly, the dissociative behavior of the cluster potentially depends on the elasticity of the target material. Broadly speaking, a cluster dissociates more easily when the target is more rigid. However, there have been very few experimental investigations of such phenomena. This is firstly because it is not easy to adjust $E_{at}$ so that it is comparable to $E_{bi}$. In addition, differences in the dissociative behavior of clusters impacting different targets may be too small to detect.

In the present study, the dissociative behavior of argon (Ar) cluster ions upon impact with metals with different elastic constants (Young's modulus) is experimentally investigated. For a cluster ion consisting of two argon atoms ($Ar_2^+$), $E_{bi}$ is about of 1 eV, which is several orders of magnitude higher than that for neutral argon clusters [13-15]. This stability of small Ar cluster ions is the result of the covalent nature of the bonds, due to the presence of unfilled electron shells. In the present study, $E_{at}$ for the Ar cluster ions was adjusted in the range 1-10 eV by varying the cluster size (number of constituent atoms), which was on the order of thousands. Five different types of metal were bombarded with cluster ions, and the mass spectra of the resulting products were measured. Most of the detected ions were $Ar_2^+$ and $Ar_3^+$, but not $Ar^+$. The branching ratio $Ar_2^+/\Sigma Ar_n^+$ ($n \geq 2$) for the ion yield, which represents the dissociation rate for the cluster ions, is plotted as a function of the impulsive stress acting at the contact between the cluster ion and the metal. This stress is calculated based on the density and Young's modulus of the cluster ions and the metal targets. We found that the branching ratio increases with the impulsive stress for the metal. Based on the results obtained in this study, a new method is proposed for evaluating the physical property of materials.

Details of the experimental setup have been previously reported [16, 17] and they are summarized here. Neutral Ar clusters were generated by supersonic expansion of Ar gas at a pressure of 1.2 MPa through a nozzle with a diameter of 0.1 mm. The neutral clusters were ionized by electron impact from a tungsten filament, and the resulting cluster ions were then extracted at an acceleration voltage of 2~10 kV. At this stage, the cluster size ranged from 1 (monatomic) to ~7000 atoms/cluster. Monatomic ions were removed from the cluster ion beam using a magnet. Furthermore, the required cluster size was selected by employing a time-of-flight technique using two pairs of ion deflectors. The flight time of a cluster ion between the two pairs of ion deflectors depends on the cluster size. Therefore, clusters with a specific size can be extracted by adjusting the delay time between the two pairs of deflectors. In the present experiment, the peak cluster size N could be selected in the range 500 to 5000 atoms/cluster, and a cluster size resolution (N/ΔN) of more than 10 was achieved by shortening the pulse width at the ion deflectors. As a result, $E_{at}$ could be varied in the range 0.5-20 eV with an energy width of about 15%. It should be emphasized that such controllability made it possible to detect differences in the dissociative behavior of cluster ions impacting different kinds of metal targets. The size-selected cluster ion beam (beam diameter; ~0.5 mm) was used to bombard the target at an incident angle of 45° under a pressure of

$3\times10^{-6}$ Pa at room temperature. The mass spectra of ions emitted from the target, which included dissociated Ar cluster ions and sputtered ions, were measured using a reflectron-type mass analyzer with a flight tube length of 1.8 m (KNTOF-1800-SL203, Toyama, Kanagawa, Japan). Its mass resolution (m/Δm) was approximately 200 for m/z = 700.

The samples were polycrystalline plates of silver, copper, gold, platinum, and tungsten, each with dimensions of $10\times10\times0.5$ mm$^3$ (Nilaco Co., Ltd.). The samples were ultrasonically cleaned in methanol before being placed in the ion bombardment chamber. The sample holder was made from stainless steel, and clasped the sample using a 1-mm-wide border region of its front surface, in addition to covering its entire back surface. For practical purposes, since an Ar cluster ion containing even several thousand atoms has a diameter of only a few nanometers [18], the sample size can be regarded as infinite. Accordingly, it is likely that elastic waves caused by ion impact freely propagate through the metal.

Figure 1(a) shows the mass spectra of ions detected from a silver sample under bombardment with cluster ions containing 1500 Ar atoms ($Ar_{1500}^+$) at an acceleration voltage of 5 kV ($E_{at}$ = 3.3 eV). The vertical axis is the ion yield, which represents the detected ion count divided by the incident ion current. In addition to those due to dissociated Ar cluster ions, many additional peaks are observed in the spectrum, especially in the low-mass region (m/z ≤ 100). Most are associated with contamination on the metal surface, which was sputtered away during cluster ion bombardment. Some peaks can also be assigned to fragment ions ($SiC_3H_9^+$) of poly-dimethylsiloxane (PDMS) and to silver sulfide ions ($Ag_2S^+$). PDMS is often detected when carrying out surface analysis using techniques such as secondary ion mass spectrometry [19], and its origin is thought to include the pumping oil and the resin in the sample case. It is well known that a sulfide layer can easily form on silver surfaces in atmosphere [20]. Such contaminating species should be removed before measurements are carried out on the dissociation of cluster ions, because some collision energy may be dissipated due to sputtering of these species. Therefore, the samples were bombarded with Ar cluster ions with higher energies ($E_{at}$ = 1-10 eV), just before carrying out measurements on dissociated ions. In the inset of Fig. 1(a), the yield of $SiC_3H_9^+$ and $Ag_2S^+$ is plotted as a function of ion dose during the higher-energy bombardment (pre-sputtering). The yield of $SiC_3H_9^+$ is seen to rapidly decrease with ion dose, while that of $Ag_2S^+$ initially increases and then decreases with ion dose. This indicates that most PDMS molecules are adsorbed on a sulfide layer. As shown in Fig. 1(b), for an ion dose of $8\times10^{13}$ ions/cm$^2$, the signals due to the contaminating species are remarkably reduced and a series of peaks associated with dissociated Ar cluster ions can be clearly observed. In addition, a periodic series of weaker signals is present between adjacent Ar cluster ion peaks. These signals can be assigned to Ar cluster ions attached to one or two water molecules. Trace amounts of water adsorbed on the inner wall of the metal nozzle may become incorporated into the Ar gas flow during cluster growth. It should be noted that almost no monatomic argon ions ($Ar^+$) were observed regardless the prominence of $Ar_2^+$ or $Ar_3^+$. The peak near m/z = 40 in Fig. 1(b) was found to be due to $K^+$ (m/z = 39) by exact assignment.

Several numerical and experimental studies have focused on the collisions between Ar clusters and different substrates [21-24]. Xu et al. carried out simulations of the scattering of Ar clusters, $Ar_n$ (n =5 to 26), from a Pt surface [21]. The clusters were found to dissociate almost completely into individual atoms at incident energies of 0.1 and 0.5 eV per atom, which is in contrast to the results obtained in the present study. This discrepancy is associated with the difference in the stability of neutral Ar clusters and Ar cluster ions. Ikegami et al. investigated the most stable structures of Ar cluster ions, $Ar_n^+$ (n=3-27), using quantum mechanical calculations [25]. They reported that the charge of an Ar cluster ion is localized on the three central atoms, which form a trimeric ion core, and this core is much more stable than the shell of neutral argon atoms surrounding it. As a result, the energy corresponding to the dissociation of $Ar_3^+$ to $Ar_2^+$ + Ar, and of $Ar_2^+$ to $Ar^+$ + Ar, is 0.2 and 1.2 eV, respectively. Such a core-shell structure could also be present in large cluster ions consisting of thousands of Ar atoms. Consequently, the Ar cluster ions in the present study dissociate into many neutral Ar atoms and small cluster ions such as $Ar_3^+$ or $Ar_2^+$. Based on these considerations, the branching ratio for the dissociated ion yield, $\eta_{di} = Ar_2^+ / \Sigma Ar_n^+$ (n ≥ 2), obtained from the mass spectra, is used as an index representing the rate of dissociation of cluster ions.

In the inset of Fig. 1(b), $\eta_{di}$ is plotted as a function of ion dose during the pre-sputtering process. It can be seen that $\eta_{di}$ rapidly increases initially, and then decreases and saturates. This behavior is similar to that for the contaminating ions, as shown in the inset of Fig. 1(a). Similar experiments were carried out for the other metal samples. It was found that in all cases, $\eta_{di}$ saturated for a pre-sputtering ion dose of greater than $5 \times 10^{13}$ ions /cm$^2$, although at low ion doses, the behavior varied from metal to metal. From these results, pre-sputtering is found to effectively remove contaminating species and stabilize $\eta_{di}$ for the metals. Consequently, pre-sputtering at a dose of $1 \times 10^{14}$ ions /cm$^2$ was routinely applied to all samples. It should be noted that the change in $\eta_{di}$ is very sensitive to a very thin layer of the samples.

The mass spectra of dissociated ions produced during bombardment of the silver surface with cluster ions with $E_{at}$ varying from 0.5 to 15 eV/atom were next measured. Here, adjustment of $E_{at}$ was carried out by altering cluster size at acceleration voltage of 5.0 and 7.5 kV. As $E_{at}$ increased, a shift towards smaller ions was observed in the size distribution of the dissociated components. Furthermore, when $E_{at}$ was greater than 10 eV/atom, sputtered $Ag^+$ ions began to be observed. Figure 2 shows the dependence of $\eta_{di}$ on the velocity of the incident cluster ions, which is calculated as $(2E_{at}/m)^{1/2}$, where m is the mass of the cluster ion. It can be seen that $\eta_{di}$ approaches zero or one at low or high velocity, respectively. It is noteworthy that $\eta_{di}$ shows almost a linear dependence on velocity in the range 4-6 km/s. As will be shown later, this simplifies the analysis of collisions between the Ar cluster ions and metals. In Fig. 2, it can also be seen that $\eta_{di}$ at 7.5 kV is slightly lower than that at 5.0 kV. This is thought to result from a buffering effect, in which the shell of the larger cluster ions used at 7.5 kV softens the impact on the core ions.

Next, the mass spectra of dissociated ions for the five kinds of metal were detected by the bombardment of $Ar_{1500}^+$ at 5kV ($E_{at}$ of 3.3 eV/atom). The mass spectra are considerably different by each metal, as shown in Fig. 3. Table 1 summarizes a comparison of $\eta_{di}$ for the five kinds of metal, where the standard deviations are obtained from repeated measurements. The $\eta_{di}$ values are clearly different for each metal, even when the statistical variations are taken into account. The difference in the $\eta_{di}$ values is hardly considered to be due to electronic or chemical effect on the interaction between the Ar cluster and the metals. Accordingly, for this reason we focused on differences in the mechanical response of the metals upon impact. In a previous study, it was found that the time required for a cluster to transfer its kinetic energy to the target is significantly shorter than the deformation period of the target [11]. This indicates that the force acting between the two bodies should be regarded as an impulsive force. If it is assumed that the collision is elastic during the early stages, and that the motion is approximately linear, a compressive stress field $\sigma$ is formed at the contact region and begins to extend in opposite directions [26]. The stress propagation velocities through the cluster and the metal, $c_c$ and $c_m$, respectively, are given by

$$c_c = (E_c/\rho_c)^{1/2}, \quad c_m = (E_m/\rho_m)^{1/2}, \tag{1}$$

where $E_c$ and $E_m$ are the Young's moduli, and $\rho_c$ and $\rho_m$ are the densities of the Ar cluster ion and the metal, respectively. After a short time interval $\Delta t$, regions of the cluster and the metal with thicknesses $c_c \Delta t$ and $c_m \Delta t$, respectively, will be compressed, while remaining regions will be in an unstressed state. Note that $c_c$ and $c_m$ should be distinguished from the velocities $v_c$ and $v_m$ of the particles comprising the cluster and the metal, respectively. In the region of compressive stress, shrinkage occurs in both the cluster and the metal by an amount given by $v_c \Delta t$ and $v_m \Delta t$. respectively. Accordingly, the unit shrinkage (strain), $\varepsilon_c$ and $\varepsilon_m$, is respectively given by

$$\varepsilon_c = v_c \Delta t / c_c \Delta t = v_c / c_c, \quad \varepsilon_m = v_m \Delta t / c_m \Delta t = v_m / c_m, \tag{2}$$

where $\varepsilon_i$, $E_i$ (i = c, m) and $\sigma$ are related by Hooke's law as

$$\varepsilon_i = \sigma / E_i. \tag{3}$$

Therefore we obtain

$$v_i = \sigma c_i / E_i = \sigma (E_i/\rho_i)^{1/2} / E_i = \sigma (E_i \rho_i)^{-1/2}. \tag{4}$$

Expressing $c_i$ and $v_i$ in coordinates with respect to the center of mass of the cluster or the metal, when the two bodies are in contact,

$$V_0 - v_c = v_m \tag{5}$$

where $V_0$ is the initial velocity of the cluster ion with respect to the center of mass of the metal. From Eqs. (4) and (5), we obtain

$$\sigma = V_0 \, (E_c \rho_c)^{1/2} (E_m \rho_m)^{1/2} / \{(E_c \rho_c)^{1/2} + (E_m \rho_m)^{1/2}\}. \tag{6}$$

It should be noted that the stress in the contact region is linearly dependent on the collision velocity during the first moment of impact. In the calculation of stress for the five different metals, we used their density and Young's modulus shown in Table1 [27]. For the Ar cluster ions, we calculated their densities by assuming their shape are spherical and using values for their diameter, which was calculated by Xiang et al [18]. The calculated density becomes close to a value of solid Ar, as a cluster size increases over 500 atoms per cluster. Accordingly, we used a value of solid Ar at a temperature of 40 K ($1.7 \times 10^3$ kg/m$^3$) [27] for the density, together with the Young's modulus of solid Ar at 82 K (2.38 GPa), which was determined by the measurement of Brillouin scattering by Gewurtz and Stoicheff [28]. Figure 4 shows the relationship between $\eta_{di}$ and the calculated stress for the different metals, and it is seen to be close to linear. Except for the case of Au and Cu, $\eta_{di}$ increases with Young's modulus. The exception for these two metals is because the density of Au is higher than that of Cu. The linear relationship seen in Fig. 4 is reasonable since both $\eta_{di}$ and the stress depend linearly on the collision velocity, as indicated in Fig. 2 and Eq. (6), respectively. The model mentioned above didn't consider an influence by the energy dissipation through the viscosity or friction between the cluster and metal. However, it shows the validity of the model that magnitude correlation in the stress well corresponded with that in $\eta_{di}$.

In summary, we have investigated the dissociation of Ar cluster ions when they collide with several kinds of metals at $E_{at}$ comparable to $E_{bi}$. Mass spectrometric analysis of the dissociated ions has just become possible by employing the pre-sputtering cleaning of the metal surface. The branching ratio of dissociated ions $\eta_{di}$, representing the dissociation rate, is significantly different for each metal. Analysis assuming an elastic theory concludes that difference in $\eta_{di}$ is brought about by the difference in mechanical response to the impact and that $\eta_{di}$ increases linearly with impulsive stress acting on the contact area between the cluster and the metal.

 The significance of the present study is the ability to experimentally identify a mechanical effect on nanoscale collisions. As far as we know, this is the first result showing that a collision-induced stress determines the dissociation of clusters. The result suggests that a continuum approximation used in macroscopic phenomena is also available for analysis of nanoscale collisions. Such a finding could contribute to simplification of modeling of nanoscale systems and reduction of load required for calculation. At present, it is still not clear whether the present results are peculiar to the cluster size used here. Experiments using Ar cluster ions with the size of a wider range will clarify the limit where the continuum method is effective in nanoscale systems.

 The results obtained here also suggest a new method for evaluating physical property of materials, such as density or mechanical strength. Mechanical strength like Young's modulus is normally evaluated by measuring the amount of the deformation under a known load, during which the sample needed to be held very precisely [29]. This becomes increasingly difficult as materials acquire nanometer dimensions. However, the use of a cluster ion beam is applicable to any sample regardless of its shape or size (down to the beam size). As described earlier, $\eta_{di}$ is sensitive to a very thin layer of the sample. Thus, a method using an Ar cluster ion beam is applied to an advanced measurement such as 3D mapping of physical property of material in a real time mode, which could open a completely new area of material science.

 This work was supported by Grant-in-Aid for Scientific Research from the Japan Society for the Promotion of Science, and Development of System and Technology for Advanced Measurement and Analysis, Japan Science and Technology Agency. The authors would like to acknowledge Dr. Tsutomu Ikegami of Advanced Industrial Science and Technology, Japan for helpful discussions on the atomic structure of Ar cluster ions.

Captions of figures and table

FIG. 1.   Mass spectra of ions detected from silver sample during bombardment with $Ar_{1500}^+$ at 3.3 eV per atom, before (a) and after (b) higher-energy (1-10 eV) ion bombardment (pre-sputtering) at ion dose of $8\times10^{13}$ ions /cm$^2$. The inset in (a) plots the yields of $SiC_3H_9^+$ and $Ag_2S^+$ as a function of the dose of higher-energy ions. The inset in (b) plots $\eta_{di}$ ($Ar_2^+/ \Sigma Ar_n^+$ ($n \geq 2$)) as a function of the dose of higher-energy ions.

FIG. 2.   Dependence of $\eta_{di}$ on velocity of incident cluster ions, calculated as $(2E_{at}/m)^{1/2}$, where m is the cluster ion mass. An almost linear relationship exists in the velocity range 4-6 km/s. At low and high velocities, $\eta_{di}$ approaches zero and one, respectively. $\eta_{di}$ at 7.5 kV is slightly lower than that at 5.0 kV.

FIG. 3.   Mass spectra of dissociated ions detected from five kinds of metal, silver, gold, cupper, platinum, and tungsten sample by the bombardment with $Ar_{1500}^+$ at 5 keV ($E_{at}$: 3.3 eV).

Table 1.   Values of $\eta_{di}$ for dissociated cluster ions during bombardment with $Ar_{1500}^+$ at 5 kV ($E_{at}$ = 3.3 eV) for five kinds of metals. The mean values and standard deviations were obtained by repeated measurements on each sample. The density and Young's modulus for the metals used for calculation of the impulsive stress are taken from the literature [27].

FIG. 4.   Relationship between $\eta_{di}$ and stress in contact region between cluster and metal. The $\eta_{di}$ value depends on the type of metal and has an almost linear dependence on stress. This is reasonable since both $\eta_{di}$ and the stress depend linearly on the collision velocity.

FIG. 1 (color on line).

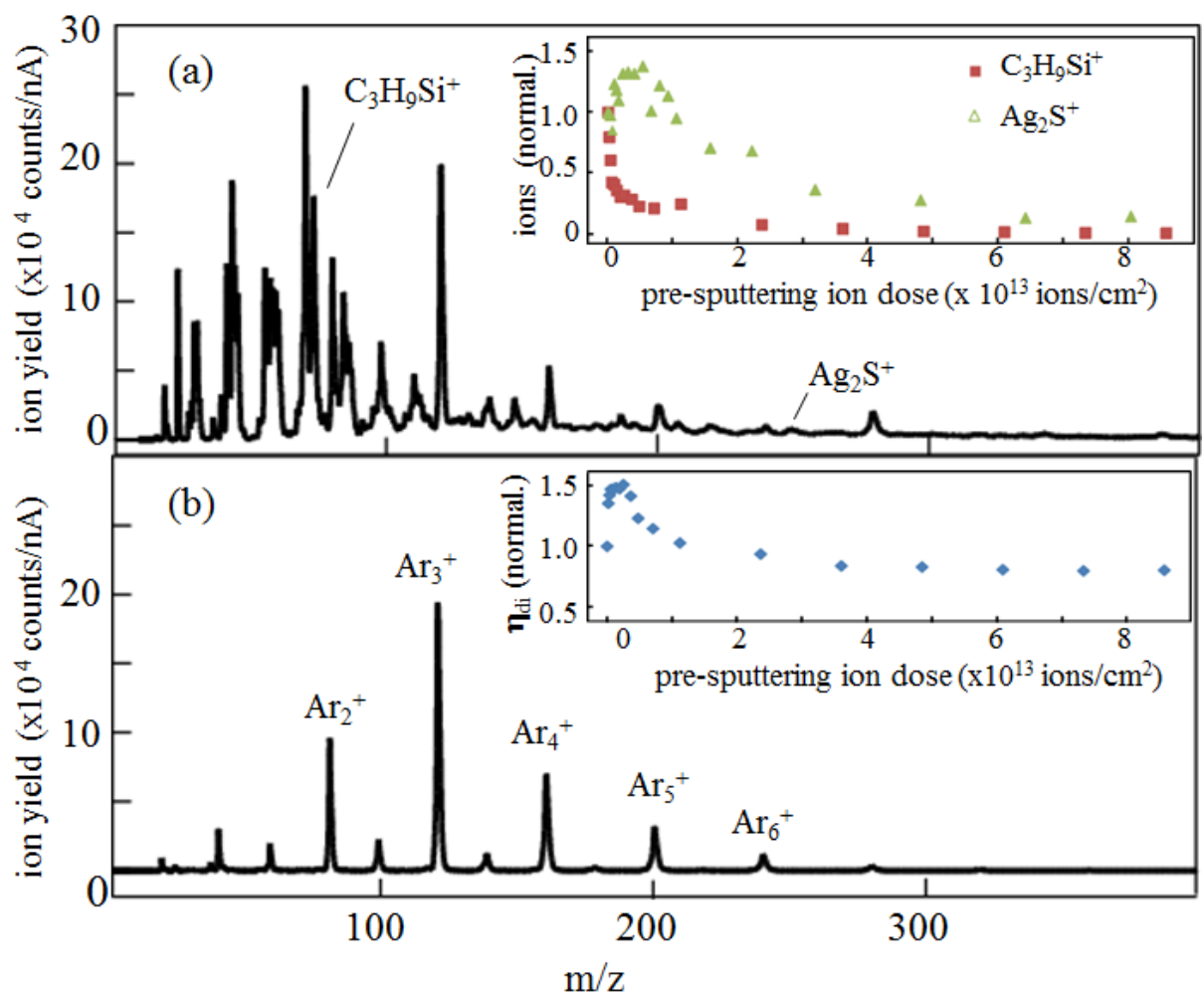

FIG. 2 (color on line).

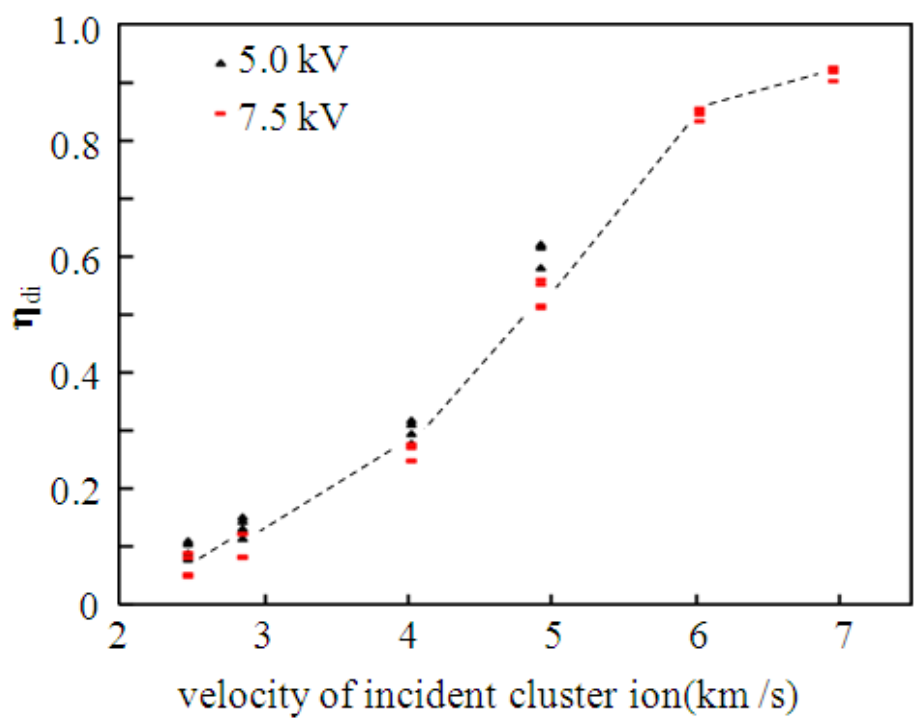

FIG. 3

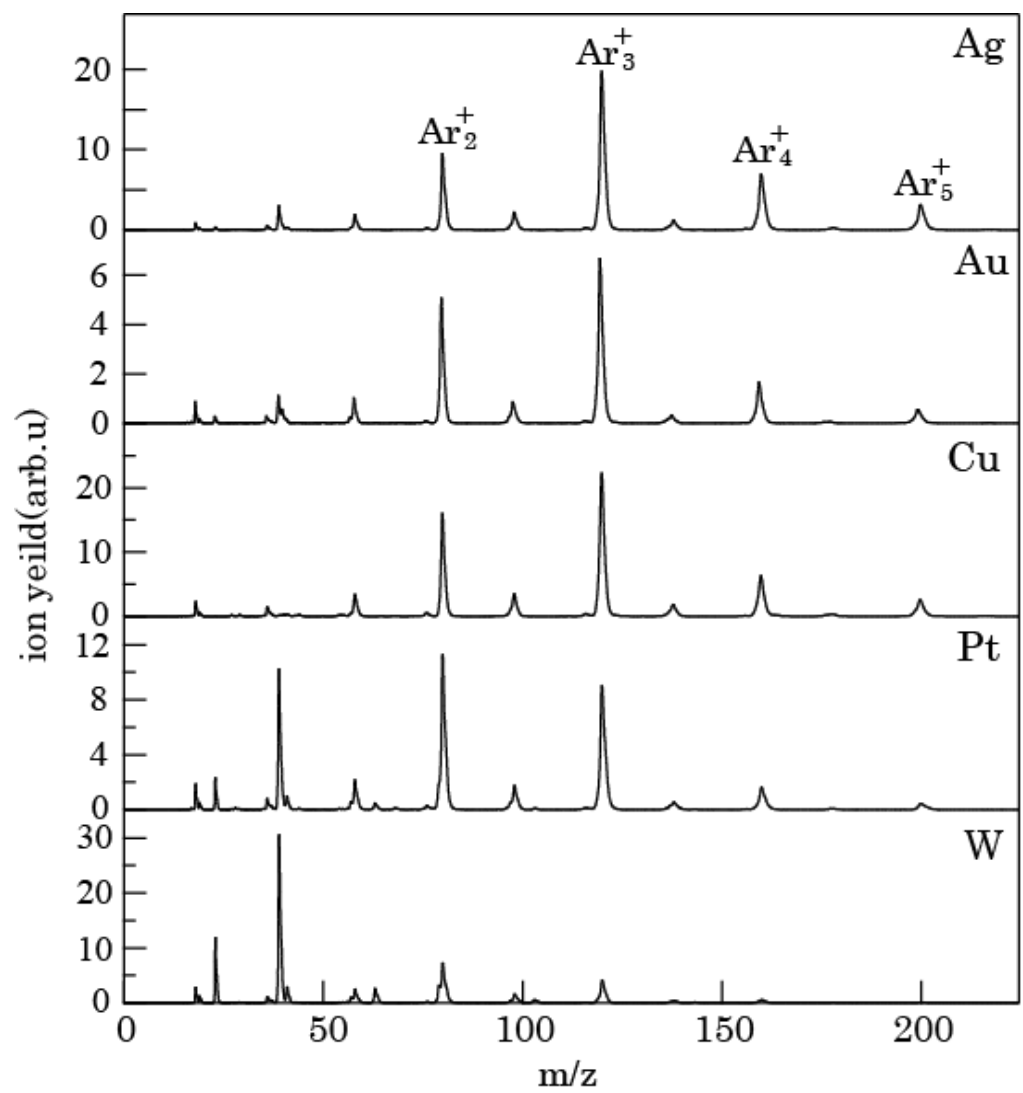

Table 1.

| metal | $\eta_{di}$ (mean ± σ) | density (× $10^3$ kg/$m^3$) | Young's modulus (GPa) |
|---|---|---|---|
| Ag | 0.23 (± 0.015) | 10.5 | 73 |
| Au | 0.39 (± 0.010) | 19.3 | 79 |
| Cu | 0.33 (± 0.002) | 8.9 | 110 |
| Pt | 0.48 (± 0.022) | 21.5 | 168 |
| W | 0.59 (± 0.014) | 19.3 | 410 |

FIG. 4

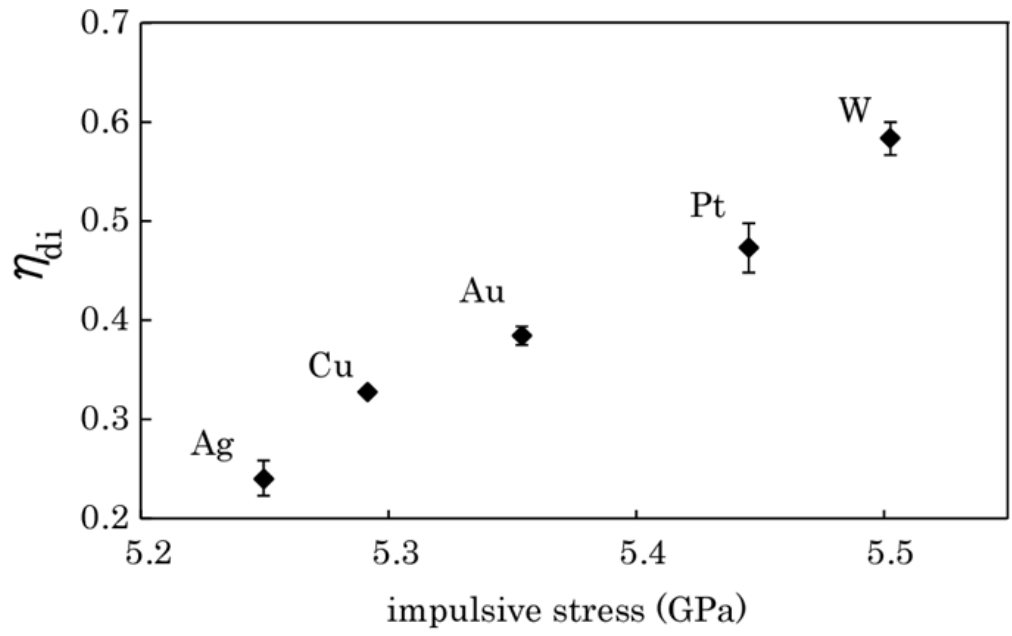